\documentstyle[prd,aps,epsf]{revtex}
\begin{document}
\draft

\title{Initial data for superposed rotating black holes}
\author{William Krivan}
\address{Department of Physics, University of Utah,
         Salt Lake City, UT 84112 \\
    and Institut f\"ur Astronomie und Astrophysik, Universit\"at 
        T\"ubingen, D-72076 T\"ubingen, Germany}
\author{Richard H. Price}
\address{
Department of Physics, University of Utah, Salt Lake City, UT 84112}
\maketitle
\begin{abstract}
  The standard approach to initial data for both analytic and
  numerical computations of black hole collisions has been to use
  conformally-flat initial geometry. Among other advantages, this
  choice allows the simple superposition of holes with arbitrary mass,
  location and spin. The conformally flat restriction, however, is
  inappropriate to the study of Kerr holes, for which the standard
  constant-time slice is not conformally flat.  Here we point out that
  for axisymmetric arrangements of rotating holes, a nonconformally
  flat form of the 3-geometry can be chosen which allows fairly simple
  superposition of Kerr holes with arbitrary mass and spin. We present
  initial data solutions representing locally Kerr holes at large
  separation, and representing rotating holes close enough so that
  outside a common horizon the spacetime geometry is a perturbation of
  a single Kerr hole.
\end{abstract}


\section*{I. Introduction and overview}
The study of black hole collisions has been of recent interest in
connection with the development of gravitational wave
detectors\cite{fh}, and as a proving ground for numerical
relativity\cite{grandchallenge}. In addition, it will give
unprecedented insights into the nonlinear workings of relativistic
gravitation in strong field situations.

The approach in numerical work has usually been to use the
prescription of Bowen and York\cite{BowenYork} to generate initial
data representing the holes.  In this prescription, the initial value
problem is greatly simplified by the assumption that the initial
geometry is conformally flat. This approach, though, has some
disadvantages for studying collisions with net angular momentum, since
the final hole formed will be a Kerr hole, and --- at least for
standard slicings --- the spatial geometry of a Kerr hole is not
conformally flat.  This means that computations of radiation from
coalescence will contain early radiation due to the relaxation of
Bowen-York holes to Kerr holes, as well as radiation more directly
attributable to the coalescence itself.  Still, this represents an
astrophysical imperfection rather than a major obstacle to numerical
work. For analytic work, however, it is useful to consider initial
configurations close to the final stationary hole so that the initial
configuration, and the evolution, can be treated with perturbation
theory\cite{AP1}. This technique is inherent in the methods used for
extraction of radiation \cite{extraction} in full numerical
relativity.  First and second-order perturbation theory have also been
remarkably successful in the ``close limit approximation,'' the
analysis of collisions starting from small
separations\cite{price_pullin94,all,GlNiPrPuprl,GlNiPrPuby,GlNiPrPuprd,GlNiPrPucqg,boost1,boost2,Gl,Pu}.

In this paper we present an approach to the generation of initial data
that fits the needs of perturbative analysis. Since this approach has
essentially the same computational simplicity as that in the
conformally flat approach, it is our hope that these solutions will be
considered as starting points for numerical relativity evolution. In
recent work, comparisons\cite{all,boost2} between analytic and
numerical results have been very helpful to both efforts.

Within the context of a more general formalism, we will present two
families of initial data solutions. The first family of solutions will
represent an axisymmetric superposition of two equal mass Kerr holes,
parameterized by the mass of each hole and the spin of each hole, and
a measure of the separation between them. The solution will have the
crucial feature that the holes have local ``Kerricity.'' That is, in a
region around each hole small compared to the spacing between the
holes, the 3-geometry and extrinsic curvature will have the form of
the Kerr initial data, aside from corrections that vanish in the limit
of infinite separation.
The specific solution given serves to illustrate a method of
superposition which can be generalized in an obvious way to a large
number of axisymmetric initial configurations. In particular, it can
be used to construct initial data for unequal (different mass,
different spin) Kerr holes, for more than two holes, for a collection
of Kerr and Bowen-York spinning holes, or for a collection of
Bowen-York holes (although in this last case the method involves only
conformally flat geometries and is equivalent to the Bowen-York
method).  For any solution constructed by this method the hole will
locally have the same form as if it were isolated.

A very important feature of these solutions is their close limit. If
we reduce the separation between two holes, an initial horizon will
surround both holes, but in the limit of small separation the solution
outside that horizon is not a single Kerr hole.  This is very
different from what happens for nonrotating holes, and this difference
is one of the major points we wish to make in this paper.  The fact
that the Kerr geometry is not the ``natural'' close limit for a
sequence of initial data sets for two locally Kerr holes is not a flaw
in our superposition method, but is to be expected for general initial
data sets.

The fact that the close limit of rotating holes is not a Kerr hole is
surprising because it is so very different from the Schwarzschild
case. For initial value solutions for two nonspinning holes, such as
the solutions of Misner\cite{Mi} or of Brill-Lindquist\cite{BrLi}, if
the separation between the holes is reduced one expects an
all-encompassing horizon to exist on the initial hypersurface, and for
the initial value solution outside that initial horizon to approach
the Schwarzschild geometry. And this is just what
happens\cite{price_pullin94,AP2}. In the mathematics, as in intuition,
the only source of nonsphericity is the separation of the initial
holes, and as that separation is reduced, the solution outside the
event horizon becomes spherical, and hence becomes Schwarzschild
initial data.

Crucial to the simplicity of this process is the fact that the
Schwarzschild initial data is singled out by the requirement of
sphericity. The Kerr initial solution, by contrast, requires much
more: a set of relationships involving the initial extrinsic curvature
as well as the initial 3-geometry. If these relationships are
satisfied for the individual holes, there is no reason for them to be
satisfied by the superposition of two holes. One can see a rough
quantitative sign of this in the following. For a rotating hole of
mass $M$ and angular momentum $J$, the quadrupole moment in the
3-geometry has a magnitude of order $J^2/M$. If we add two such holes
we expect to infer from the distant geometry that we have
approximately twice the mass, twice the the angular momentum, and
(from the $1/r^2$ part of $g_{00}$) twice the quadrupole moment. The
quadruple moment will therefore be too small by a factor of two for
the result to be a Kerr hole.  In a manner of speaking, then, what
makes two Kerr holes different from a single hole is not simply the
separation, it is the more basic fact that there are two of them.
There is a completely different point of view from which this can be
understood. The Kerr solution is singled out as a spacetime geometry
with a stationary horizon. The initial value solution --- the
3-geometry and extrinsic curvature --- are not special as initial
value solutions. By contrast, the initial value solutions for
Schwarzschild (spherical 3-geometry, zero extrinsic curvature) is
special.

We are pointing out here that the Kerr close limit should not be
expected from the superposition of two Kerr holes. This should not be
confused with the fact that the Kerr geometry is the natural late time
result for the {\em evolution} of two holes. Due to the evolution, the
geometry and extrinsic curvature on a late hypersurface, outside the
final horizon, will have approximately Kerr form.  For that reason, it
is useful to have  solutions available with a Kerr close limit.

In Sec.\,III  below we construct such a solution. This solution is
somewhat contrived, but any other multihole solution with a Kerr close
limit will also be contrived (except of course, the solution produced
by evolution). Our particular contrived solution has the unappealing
feature that there is a singularity where there should not be a
physical reason for a singularity.  Since this contrived solution is
meant to be used only for small separations, in which case this
singularity will be well inside the horizon surrounding the two
initial holes.  In judging the importance of this singularity, it is
necessary to keep in mind the use to which the close-limit solution is
to be applied. For this initial value solution (or any other initial
value solution with a Kerr close limit) it is possible to use
close-limit perturbation theory to find the outgoing radiation
generated in the evolution of the initial data. If these initial data
are used for numerical relativity, very useful comparisons with
analytic results are possible. For such comparisons the unwanted
singularity is of concern only if it causes problems in the numerical
method. For numerical relativity computations that use horizon
excision the singularity should be of no concern whatever.

An alternative approach has recently been given by Baker and
Puzio\cite{bakerpuzio} to the construction of initial data sets for
axially symmetric non-conformally flat, initial data.  That method
starts with a specification of the 3-geometry and solves for partial
information about the extrinsic curvature. The method has the
advantage, in principle, of allowing generalization to a broader class
of axisymmetric solutions, in particular to rotating holes which are
initially moving. The method has the disadvantage that it is more
difficult to implement, and has not yet been tested numerically.

The remainder of the paper is organized as follows: In Sec.\ II we
present the basic method of superposition, and the numerical solution
for two identical ``locally Kerr'' holes, including a numerical
demonstration of local Kerricity. In Sec.\ III we show that the locally 
Kerr solutions of Sec.\ II do not have a Kerr close limit, and we
introduce solutions that do have a Kerr close limit.  A discussion and 
conclusions are presented in Sec.\ IV.

\section*{II. Mathematical Approach}

We start with the Brill-wave\cite{BrillWave} form of the metric, as
studied by Brandt and Seidel \cite{BS1,BS2,BS3}. In place of their
radial coordinate $\eta$ which is appropriate to the description of
both sides of a throat, we use the alternative coordinate $\bar{r}$,
related to $\eta$ by $d\eta=d\bar{r}/\bar{r}$. In terms of this coordinate, and
the Brandt-Seidel notation, the form of the metric is
\begin{equation}
  \label{genmet}
ds^2=\Phi^4\left[e^{-2q}\left(d\bar{r}^2
+\bar{r}^2d\theta^2\right)
+\bar{r}^2\sin^2\theta\,d\phi^2
\right]\ ,
\end{equation}
where the $q$ and the conformal function $\Phi$ are functions of
$\bar{r}$ and $\theta$.
The extrinsic curvature has only $\bar{r}\phi$ and $\theta\phi$ components,
and in the Brill-Brandt-Seidel notation those are denoted in terms of 
$\widehat{H}_E$  and $\widehat{H}_F$, conformally related  
functions of $\bar{r},\theta$ as 
\begin{equation}
  \label{extrin_r_1}
K_{\bar{r}\phi}=\bar{r}^{-2}\Phi^{-2}\widehat{H}_{E}\sin^2\theta \; ,
\hspace*{.55in} 
K_{\theta\phi}=\bar{r}^{-1}\Phi^{-2}\widehat{H}_{F}\sin\theta\ .
\end{equation}
The initial value equations of Einstein's theory impose no constraints on
$q$, but the momentum constraint requires that the $\widehat{H}$s satisfy
\begin{equation}
  \label{momcon}
  \bar{r}\partial_{\bar{r}} (\widehat{H}_E)\sin^3\theta
+\partial_\theta(\widehat{H}_F\sin^2\theta)=0\ .
\end{equation}
The fact that in this form the momentum constraint is linear, and does not
depend on the spatial geometry (i.e., is independent of both $q$ and $\Phi$)
is what ultimately will allow the fairly simple superposition to be 
described below.

Once $q$, and $\widehat{H}$s solving (\ref{momcon}), are chosen, one finds the
conformal function $\Phi$ by solving the Hamiltonian constraint
\begin{equation}
  \label{hamconst}
{\cal L}_{\rm ham}(\Phi)=
 -\Phi^{-7}\left[\widehat{H}_E^2\sin^2\theta+\widehat{H}_F^2\right]/
\left(
4\bar{r}^6
\right)\ ,
\end{equation}
where
\begin{equation}
  \label{lindef}
  {\cal L}_{\rm ham}(\Phi)\equiv
  \left\{\frac{1}{\bar{r}^2}\frac{\partial}{\partial\bar{r}}
\left(\bar{r}^2\frac{\partial}{\partial\bar{r}}\Phi\right)  
+\frac{1}{\bar{r}^2\sin{\theta}}\frac{\partial}{\partial\theta}
\left(
\sin\theta\frac{\partial\Phi}{\partial\theta}
\right)
\right\}
-\frac{\Phi}{4\bar{r}^2}\left[\bar{r}\frac{\partial}{\partial\bar{r}}
\left(\bar{r}\frac{\partial}{\partial\bar{r}}q\right)
+\frac{\partial^2}{\partial\theta^2}q
\right] \ .
\end{equation}
Note that the terms in curly brackets have  the form of the flat space
Laplacian.

For a Kerr hole of mass $M$ and angular momentum parameter (angular
momentum per mass) $a$, the forms of the functions are
\begin{equation}
  \label{qKerr}
  e^{-2q_{K}}\equiv  \frac{
r^2+a^2\cos^2\theta
}{
r^2+a^2+\frac{2Ma^2r\sin^2\theta}{r^2+a^2\cos^2\theta}
}\ ,
\end{equation}
\begin{equation}
  \label{HEKerr}
  \widehat{H}_{EK}=\frac{aM
\left[
(r^2-a^2)(r^2+a^2\cos^2\theta)+2r^2(r^2+a^2)
\right]
}{
\left[r^2+a^2\cos^2\theta\right]^2
}\ ,
\end{equation}
\begin{equation}
  \label{HFKerr}
  \widehat{H}_{FK}=\frac{
-2a^3Mr\left(r^2-2Mr+a^2
\right)^{1/2}\cos\theta \sin^2\theta
}{
\left[r^2+a^2\cos^2\theta\right]^2
}\ ,
\end{equation}
\begin{equation}
  \label{KerrPhi}
  \Phi^4_{K}=\bar{r}^{-2}\left[r^2+a^2+\frac{2Ma^2r\sin^2\theta
}{r^2+a^2\cos^2\theta
}\right]\ .
\end{equation}
Here we have used the index ``$K$'' to denote the Kerr form, and $r$ is
a rescaling of $\bar{r}$ given by
\begin{equation}
  \label{rdef}
  r(\bar{r},M,a)\equiv\bar{r}\left(1+\frac{M+a}{2\bar{r}}\right)\!
\left(1+\frac{M-a}{2\bar{r}}\right)\ .
\end{equation}

The solution of (\ref{hamconst}) depends not only on how we choose $q,
\widehat{H}_E$ and $\widehat{H}_F$, but also on the boundary
conditions 
for $\Phi$
that are specified. To recover the Kerr solution, we must not only
specify that $q, \widehat{H}_E$ and $\widehat{H}_F$ have the forms given in
(\ref{qKerr})--(\ref{HFKerr}), we must also specify boundary
conditions that $\Phi\rightarrow1$ at
$\bar{r}\rightarrow\infty$, and that $\Phi$ has a singularity at
$\bar{r}=0$ of the form $\Phi\rightarrow \sqrt{M^2-a^2}/(2\bar{r})$.

To specify more general boundary conditions for (\ref{hamconst}), it
is useful to write
\begin{equation}
  \label{decomp}
  \Phi=\Phi_{\rm reg}+\Phi_{\rm sing}\ ,
\end{equation}
in which $\Phi_{\rm reg}$ is regular in the finite $\bar{r},\theta,\phi$
space, and $\Phi_{\rm reg}\rightarrow0$ at $\bar{r}\rightarrow\infty$. The
physical information specifying the nature of the solution to
(\ref{hamconst}) that we are seeking is put into a specific choice for
$\Phi_{\rm sing}$. Equation (\ref{hamconst}) then takes the form
\begin{equation}
  \label{hamforphireg}
  {\cal L}_{\rm ham}(\Phi_{\rm reg})=  -{\cal L}_{\rm ham}(\Phi_{\rm sing})
 -(\Phi_{\rm sing}+\Phi_{\rm reg})^{-7}
\left[\widehat{H}_E^2\sin^2\theta+\widehat{H}_F^2\right]/(4\bar{r}^6)\ .
\end{equation}
This equation can be solved iteratively for $\Phi_{\rm reg}$, and can be
viewed as a linear partial differential equation, with a known source,
at each step of the iteration.  The linear difference equations
arising at each iterative step were solved with an explicit solver
from the LAPACK package, and as a check we also solved with a
successive overrelaxation routine. A compactified radial coordinate
$r_c\equiv
\bar{r}/(\bar{r}+c)$, 
reaching a finite value at $\bar{r}\rightarrow\infty$, was used, and it
was found that values on the order of $c=1$ tended to give the best
accuracy.

It is instructive to ``find'' the Kerr geometry with the method 
just outlined. We start by taking  $q,
\widehat{H}_{E}$ and $\widehat{H}_{F}$ to be the Kerr functions given in
(\ref{qKerr}) ---(\ref{HFKerr}). Next, a choice must be 
made for 
$\Phi_{\rm sing}$. A choice  of the form
\begin{equation}
  \label{BLwithkappa}
  \Phi_{\rm sing}=1+\frac{\kappa}{2\bar{r}}\  ,
\end{equation}
will lead to a single throat solution. If, in addition, we specify
that $\kappa =\sqrt{M^2-a^2}$, we are choosing $\Phi_{\rm sing}$ to have
the singular behavior of $\Phi_K$, and the solution for $\Phi$ should
be the Kerr conformal factor. The solution for any other nonzero
choice of $\kappa$ may be considered to be a distorted Kerr hole. [If
$\kappa$ is taken to vanish, the right hand side of
(\ref{hamforphireg}) diverges, and the solution for $\Phi$ turns out
to give a nonstandard topology, of a type discussed in the next
section.]  As a test of our numerical code, we have solved
(\ref{hamforphireg}) for Kerr inputs $q_K,\widehat{H}_{EK},
\widehat{H}_{FK}$ with
the singular solution in (\ref{BLwithkappa}) and the Kerr choice
$\kappa =\sqrt{M^2-a^2}$. Numerical solutions were computed for the
case $M=1, a=0.9$, and were found to be in excellent agreement with
the analytic form $\Phi_K$ in (\ref{KerrPhi}).  For a grid with 200 
radial and 80 angular divisions, the fractional error was of order
$3\times10^{-5}$.  It should be noted that
$\Phi_K\rightarrow1+M/(2\bar{r})$ at large $\bar{r}$, whereas we have made
the choice $\Phi_{\rm sing}\rightarrow1+\sqrt{M^2-a^2}/(2\bar{r})$. The
difference in the asymptotic behaviors is supplied by $\Phi_{\rm reg}$
which must fall off as
$\Phi_{\rm reg}\rightarrow(M-\sqrt{M^2-a^2})/(2\bar{r})$.  We have
numerically verified, with an accuracy of 0.1\%, that the coefficient
of the $1/\bar{r}$ term in the computation of $\Phi_K$, is $M/2$.

As the first step in superposing Kerr holes, we take a Kerr hole, as
described above, for mass $M_1$ and spin parameter $a_1$, and 
in the following way we shift the hole to a coordinate position 
$\bar{r}=z_1, \theta=0$:
We introduce coordinates
$\bar{r}_1,\theta_1,\phi$, related to our fundamental coordinates
$\bar{r},\theta,\phi$, by
\begin{equation}
  \label{transform}
  \bar{r}_1\cos{\theta_1}=  \bar{r}\cos{\theta}-z_1 \; ,
\hspace*{.5in}  \bar{r}_1\sin{\theta_1}=  \bar{r}\sin{\theta}\ .
\end{equation}
In the new coordinates the metric has the same form as in
(\ref{genmet}) so a solution of the initial value equations is given
by (\ref{qKerr}) -- (\ref{rdef}), with $\bar{r},\theta$ replaced by
$\bar{r}_1,\theta_1$, and with the mass and angular momentum parameters
taken to be $M_1,a_1$. We now reexpress this shifted solution in terms
of the fundamental coordinates $\bar{r},\theta$; we denote the resulting
solution, with a subscript 1, as
$\Phi_1,q_1,\widehat{H}_{E1},\widehat{H}_{F1}$.
The first two functions transform 
rather simply,
and are given by
\begin{equation}
  \label{shifted_1}
  q_1(\bar{r},\theta)=q_K(\bar{r}_1,\theta_1;M_1,a_1)\; ,
\hspace*{.3in}\Phi_1 =\Phi_K(\bar{r}_1,\theta_1;M_1,a_1)\ ,
\end{equation}
where $q_K,\Phi_K$ are understood to be the functional forms of
coordinates and parameters given in (\ref{qKerr}) and (\ref{KerrPhi}).
In other words, these functions are given explicitly by
\begin{equation}
  \label{2q1}
  e^{-2q_1}\equiv  \frac{
r_1^2+\cos^2\theta_1
}{
r_1^2+a_1^2+\frac{2Ma_1^2\sin^2\theta_1}{r_1^2+a_1^2\cos^2\theta_1}
}\; ,
\end{equation}
and 
\begin{equation}
  \label{Phi1}
  \Phi^4_1=  \bar{r}_1^{-2}\left[r_1^2+a_1^2+\frac{2M_1a_1^2r\sin^2\theta_1
}{r_1^2+a_1^2\cos^2\theta_1
}\right]\ .
\end{equation}
In these expressions, the symbol $r_1$ is to be interpreted as  
\begin{equation}
  \label{r1def}
  r_1\equiv\bar{r}_1\left(1+\frac{M_1+a_1}{2\bar{r}_1}\right)\!
\left(1+\frac{M_1-a_1}{2\bar{r}_1}\right)\ ,
\end{equation}
and $\bar{r}_1,\theta_1$ are taken to be the functions of $\bar{r},\theta$
given by (\ref{transform}).

To find the shifted extrinsic curvature we simply transform the components
of the tensor $K_{ij}$ and reinterpret the results in terms of the conformally
related curvature quantities $\widehat{H}_{E1},\widehat{H}_{F1}$. This gives us
\begin{equation}
  \label{HEtrans}
 \bar{r}^{-2}
\Phi_1^{-2}\widehat{H}_{E1}\sin^2\theta
=
\frac{\partial\bar{r}_1}{\partial\bar{r}}
 \bar{r}_1^{-2}
\Phi_K^{-2}\widehat{H}_{EK}\sin^2\theta_1
+\frac{\partial\theta_1}{\partial\bar{r}}
 \bar{r}_1^{-1}
\Phi_K^{-2}\widehat{H}_{FK}\sin\theta_1  
\end{equation}
\begin{equation}
  \label{HFtrans}
 \bar{r}^{-1}
 \Phi_1^{-2}\widehat{H}_{F1}\sin\theta
=
\frac{\partial\bar{r}_1}{\partial\theta}
 \bar{r}_1^{-2}\Phi_K^{-2}\widehat{H}_{EK}\sin^2\theta_1
+\frac{\partial\theta_1}{\partial\theta}
\bar{r}_1^{-1}
\Phi_K^{-2}\widehat{H}_{FK}\sin\theta_1  \ .
\end{equation}
When these are  combined with the second relation in (\ref{shifted_1})
we find
\begin{equation}
  \label{HEtrans_fnl}
\widehat{H}_{E1}\sin^2\theta
=\left(\frac{\bar{r}}{\bar{r}_1}\right)^2
\left[\frac{\partial\bar{r}_1}{\partial\bar{r}}
 \widehat{H}_{EK}\sin^2\theta_1
+\bar{r}_1\frac{\partial\theta_1}{\partial\bar{r}}
\widehat{H}_{FK}\sin\theta_1 \right] \ ,
\end{equation}
\begin{equation}
  \label{HFtrans_fnl}
\widehat{H}_{F1}\sin\theta
=\left(\frac{\bar{r}}{\bar{r}_1}
\right)
\left[\frac{1}{\bar{r}_1}
\frac{\partial\bar{r}_1}{\partial\theta}
\widehat{H}_{EK}\sin^2\theta_1
+\frac{\partial\theta_1}{\partial\theta}
\widehat{H}_{FK}\sin\theta_1 \right] \ .
\end{equation}
In these equations, the expressions $\widehat{H}_{EK}$ and 
$\widehat{H}_{FK}$ are
understood to be the functional forms for the Kerr geometry.  That is,
they are given by (\ref{HEKerr}) and (\ref{HFKerr}) with
$\bar{r},\theta,M,a$ replaced by $\bar{r}_1,\theta_1,M_1,a_1$.  We have
used Maple to verify that the solutions in
(\ref{HEtrans_fnl}),(\ref{HFtrans_fnl}) satisfy the momentum
constraint (\ref{momcon}).

\begin{figure}
\epsfxsize=0.8\textwidth
\epsfbox{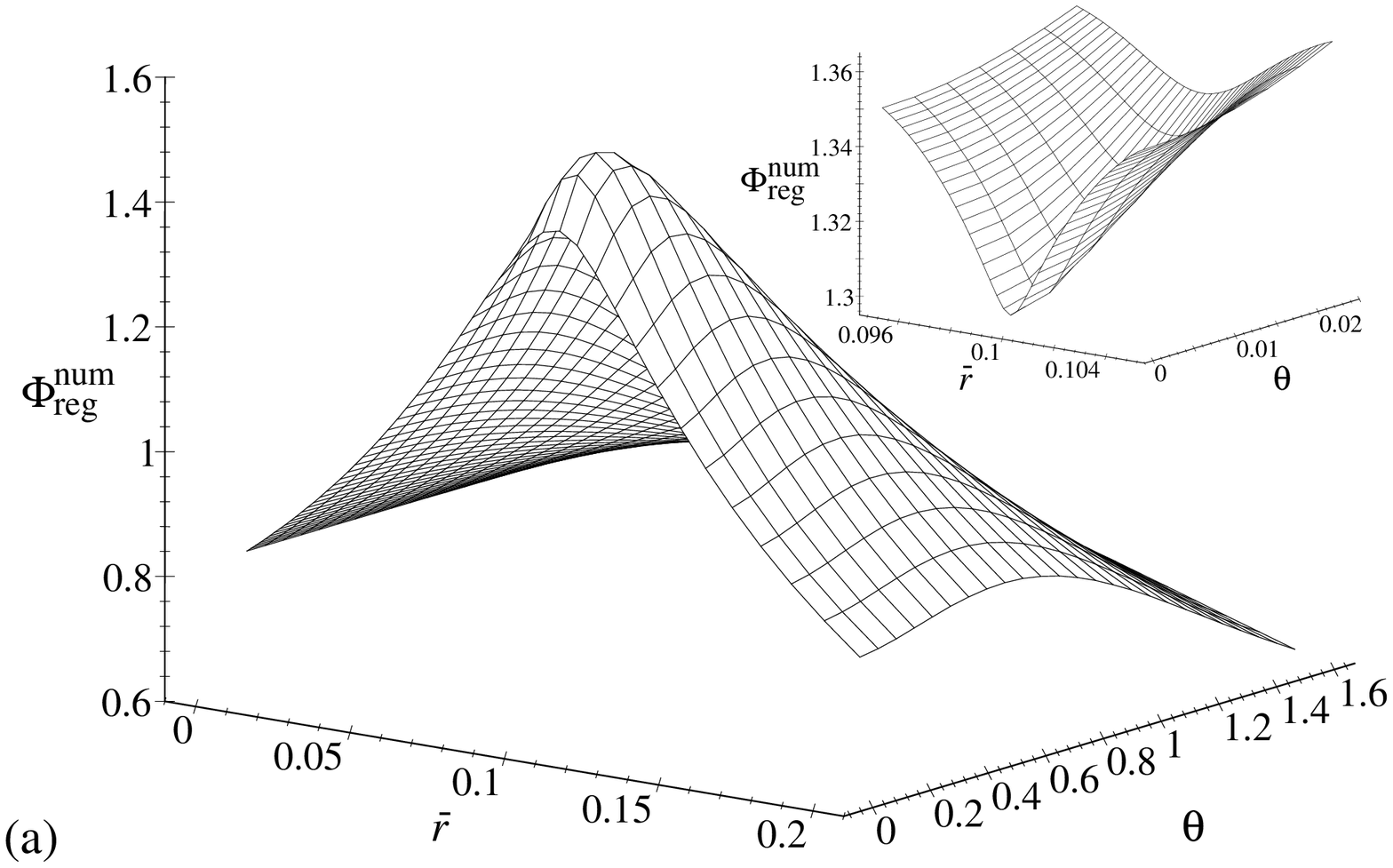}
\epsfxsize=0.8\textwidth
\epsfbox{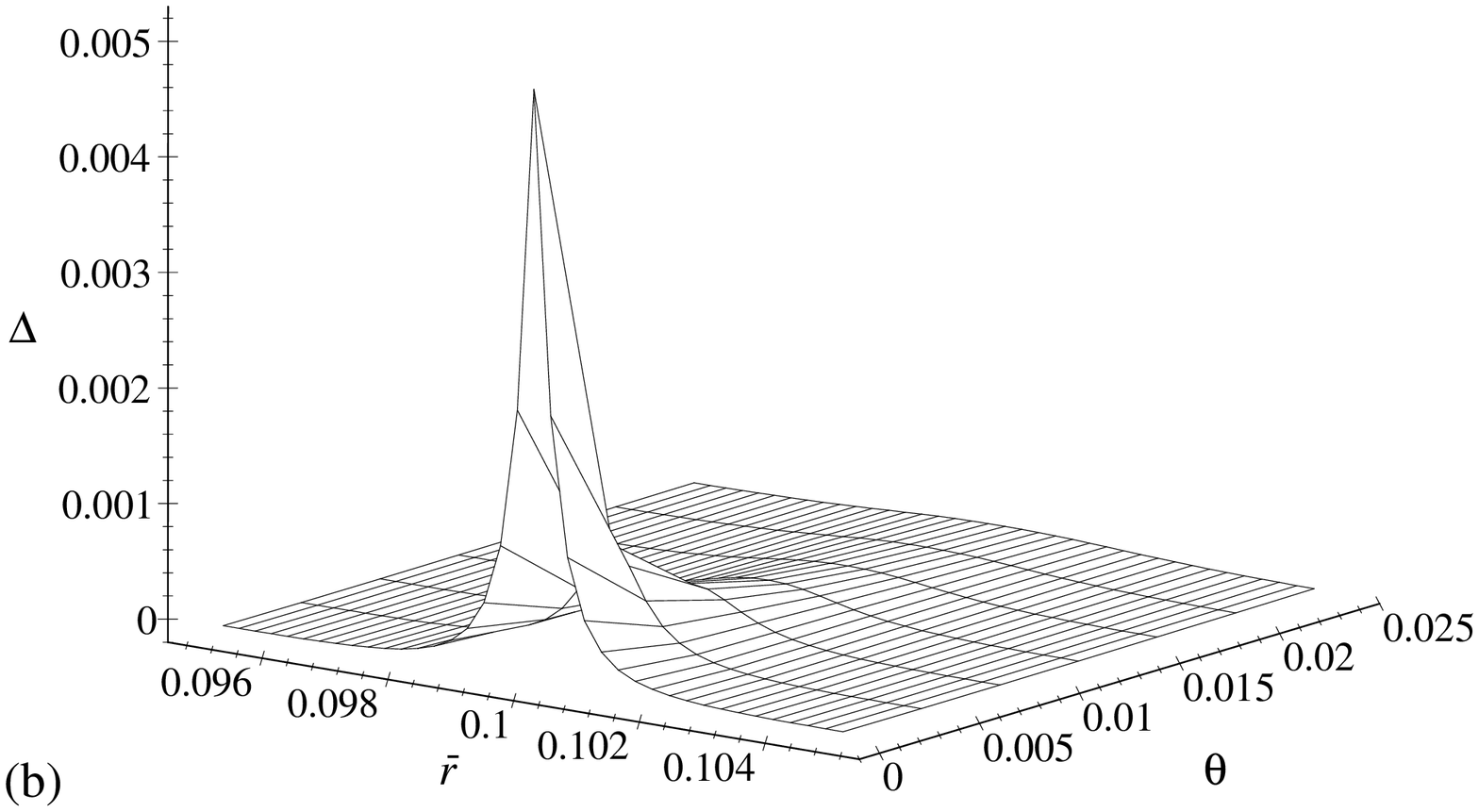}
\caption{\label{Figure-C}
(a) Numerically computed $\Phi^{\rm num}_{\rm reg}$ for one 
single Kerr black hole with $M=0.5$, $a=0.45$, located at 
$\bar{r}=z_0 = 0.1$ and $\theta=0$.
Details of the functional behavior near the singularity are displayed
in the inset.  (b) The relative difference,
$\Delta\equiv(\Phi^{\rm num}_{\rm reg}
-\Phi_{\rm reg})/\Phi_{\rm reg}$, between the numerical result
$\Phi^{\rm num}_{\rm reg}$ and the analytic expression for the 
regular part $\Phi_{\rm reg}$, for a single shifted hole. The difference 
is is shown in a neighborhood of the singularity.  }
\end{figure}

As a check on our numerical program we have used the shifted solution
of (\ref{2q1}), (\ref{HEtrans}), and (\ref{HFtrans}) in (\ref{hamforphireg})
along with the singular solution 
\begin{equation}
  \label{BLfor1hole}
  \Phi_{\rm sing}=
1+\frac{1}{2}\, \frac{\sqrt{M_1^2-a_1^2}}
                  {\sqrt{\bar{r}^2-2z_1\bar{r}\cos\theta+z_1^2}}\ .
\end{equation}
Equation (\ref{hamforphireg}) was solved numerically for $\Phi_{\rm reg}$.
The result for $\Phi_{\rm reg}$ is shown in Fig.\ \ref{Figure-C}, and is
compared with the regular part of $\Phi_1$, the analytic 
solution 
for a single shifted hole given in (\ref{Phi1}), for a hole 
with $M=0.5, J=0.225$, and
located at $z_0=0.1$.  Part (a) shows the numerically computed result 
for $\Phi^{\rm num}_{\rm reg}$ as a function of $\bar{r}$ and
$\theta$; 
an inset 
shows some detail of ``fine structure'' in the solution.  The fractional
difference between the computed and the analytic solutions
is plotted in part (b).  These computations used a numerical
grid with $1500$ radial and $800$ angular grid divisions.  The results,
both numerical and analytic, show that $\Phi_{\rm reg}$ is maximal near
the location of the throat, and --- as shown in the inset --- has some
fine scale structure where the symmetry axis passes through the
throat. The fractional difference between the computed and the analytic
$\Phi_{\rm reg}$ 
was found to exhibit second order convergence
and, at the finest grid spacing, was less than 0.5\% even near the
axis, where $\Phi_{\rm reg}$ changes on a small length scale.

In a manner similar to that just described, we can find the functions
$\Phi_2,q_2,\widehat{H}_{E2},\widehat{H}_{F2}$, describing a hole of mass $M_2$
and spin parameter $a_2$, shifted to $z_2$.  We need only replace all
``1'' subscripts in our first shifted solution by ``2.'' Since the
initial value equations are not linear we cannot  linearly
superpose the two shifted solutions, but we {\em can} superpose the
$q,\widehat{H}_E$ and $\widehat{H}_F$ functions. That is, we can take
\begin{equation}
  \label{supers}
  q=q_1+q_2\; ,\hspace*{.4in}
  \widehat{H}_{E} =\widehat{H}_{E1} +\widehat{H}_{E2}\; ,
  \hspace*{.4in}
\widehat{H}_{F} =\widehat{H}_{F1} +\widehat{H}_{F2}\ .
\end{equation}
The first of these superpositions is immediate, since $q$ is a ``free
choice'' that is not required to satisfy any constraint. The $\widehat{H}$
functions must satisfy the momentum constraint (\ref{momcon}), but
that constraint is linear and independent of the 3-geometry. Since the
first and the second shifted solution individually satisfy the linear,
geometry-independent momentum constraint, their superposition will
satisfy it. 

It remains to solve for the conformal factor $\Phi$, and this requires
specifying the singularity structure. We do this by taking
\begin{equation}
  \label{addPhis}
  \Phi_{\rm sing}-1=\left(\Phi_{{\rm sing}1}-1\right)
                        +\left(\Phi_{{\rm sing}2}-1\right)\ ,
\end{equation}
where $\Phi_{{\rm sing}1}, \Phi_{{\rm sing}2}$ are the singular parts of
$\Phi_{1}, \Phi_{2}$. This is equivalent to taking a Newtonian-like
solution
\begin{equation}
  \label{BLfor2holes}
  \Phi_{\rm sing}=
1+\frac{1}{2}\,
  \frac{\sqrt{M_1^2-a_1^2}}{\sqrt{\bar{r}^2-2z_1\bar{r}\cos\theta+z_1^2}}
+ \frac{1}{2}\,
  \frac{\sqrt{M_2^2-a_2^2}}{\sqrt{\bar{r}^2-2z_2\bar{r}\cos\theta+z_2^2}}
\  .
\end{equation}
The last step is to solve (\ref{hamforphireg}) numerically
for $\Phi_{\rm reg}$. 

With the numerical solution of the Hamiltonian constraint 
(\ref{hamforphireg}) , we
have an initial value solution for two holes, parameterized by the
masses, spins, and locations of the holes. A crucial feature of this
solution is that the spacetime is ``locally Kerr'' near the holes.
Let us consider hole 1, and let us suppose that $z_1-z_2$ is large
compared to $M_1$. Kerr-like coordinates can be introduced in terms of
which the 3-geometry and extrinsic geometry components have
approximately the form of a Kerr hole with parameters $M_1,a_1$.  At a
radial coordinate distance $r$ from hole 1, the fractional deviation
from Kerr form will be of order $r^2/|z_1-z_2|^2$, except for the
influence of hole 2 in the conformal factor. This correction will be
of order $r/|z_1-z_2|$.  The proof of these statements is
straightforward, but tedious, and the details are not given here.
Instead, we present a numerical demonstration of local Kerricity for
the case of identical holes ($M_1=M_2=0.5, J_1=J_2=0.225$)
symmetrically placed about the coordinate origin ($z_1=-z_2\equiv
z_0$).

\begin{figure}
\epsfxsize=0.8\textwidth
\epsfbox{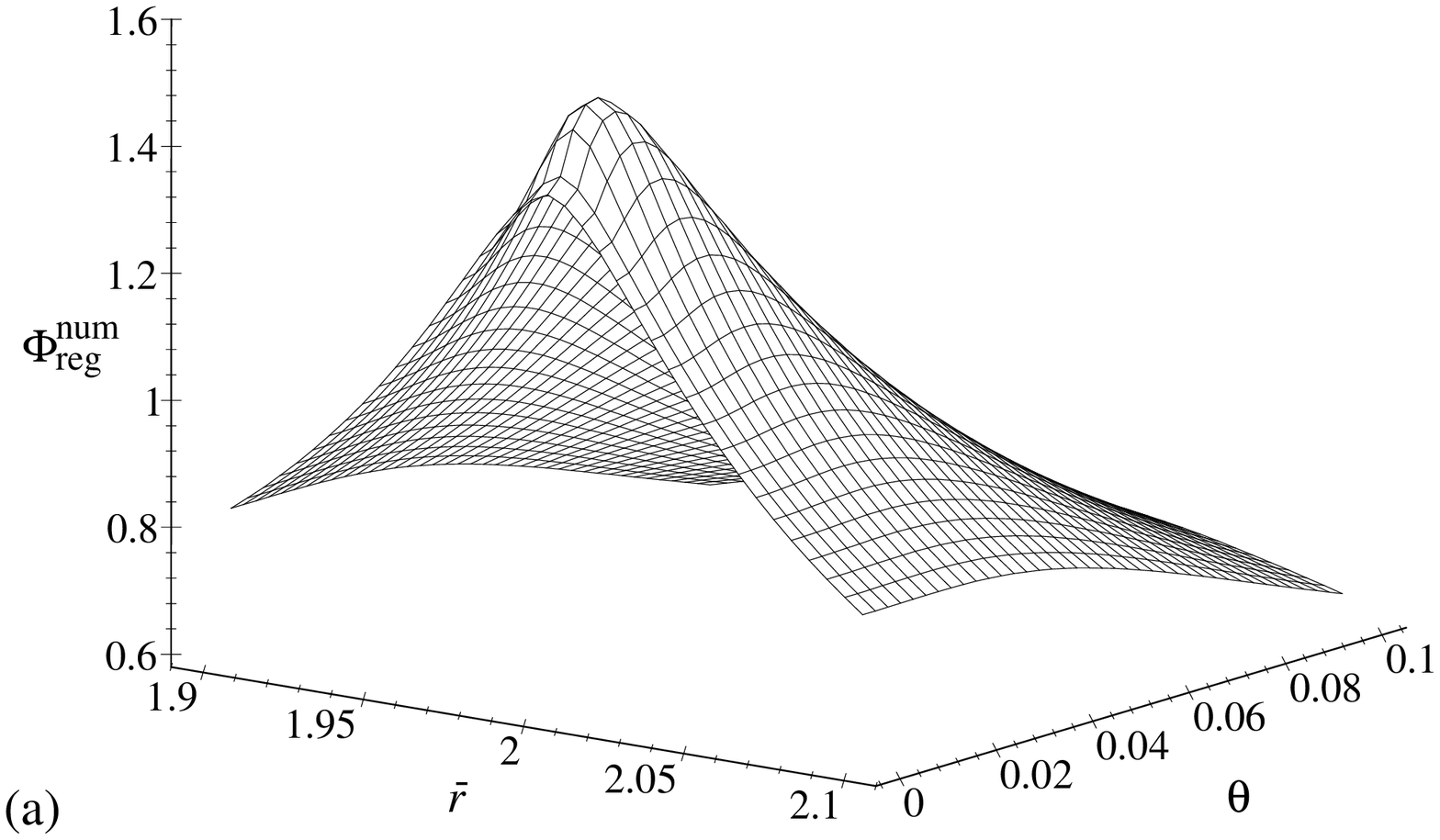}
\epsfxsize=0.8\textwidth
\epsfbox{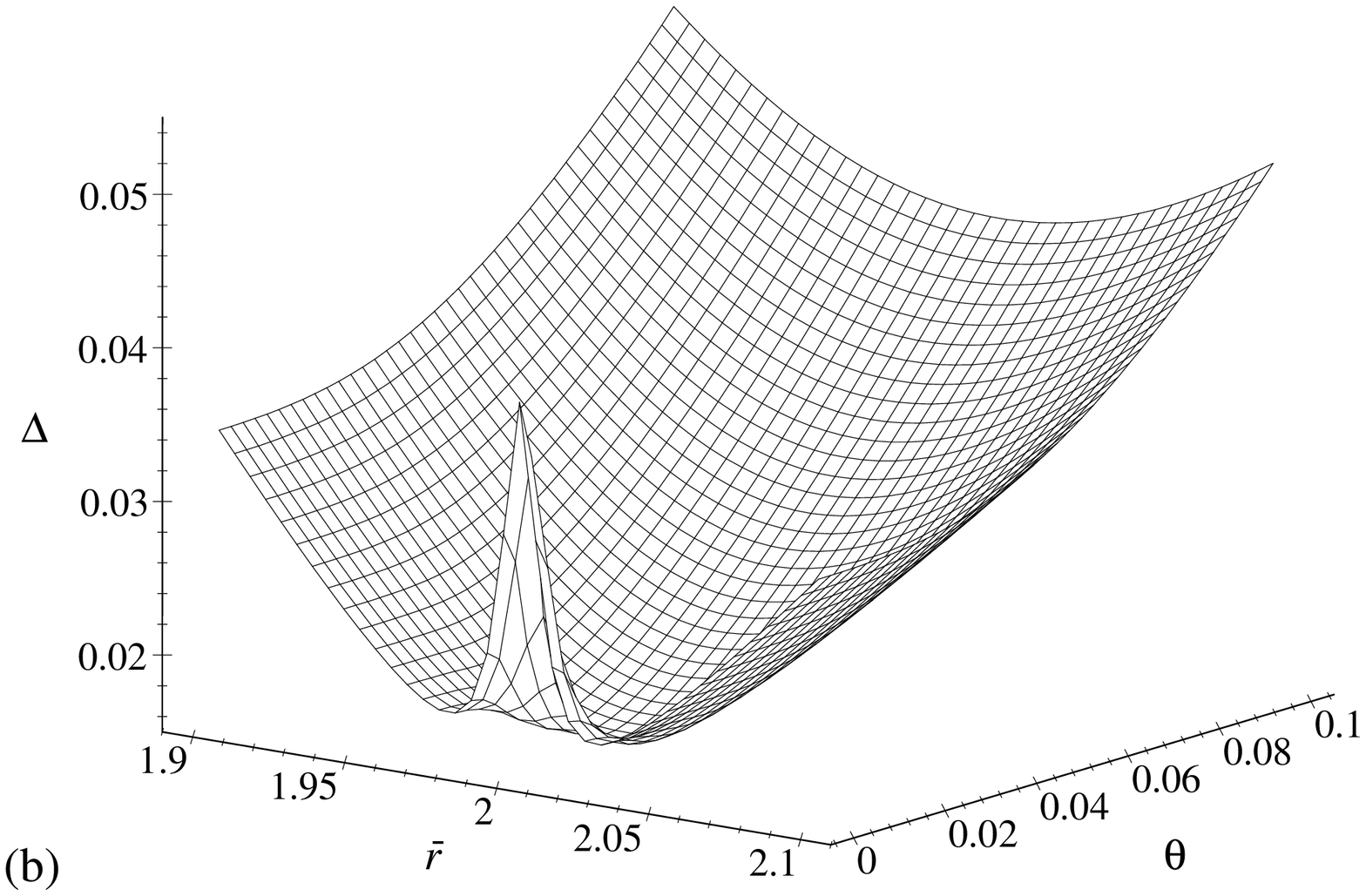}
\caption{\label{Figure-1}
(a) $\Phi^{\rm num}_{\rm reg}$, computed for two black holes with masses 
$M_1 = M_2
= 0.5$ and angular momenta $J_1 = J_2 = 0.225$, separated by a
coordinate distance $\bar{r} = 2 z_0 = 4.0$.  (b) The relative
difference, $\Delta\equiv ({\Phi_{\rm reg}^{\rm num}}'
-\Phi_{\rm reg})/\Phi_{\rm reg}$, between the
computed function $\Phi^{\rm num}_{\rm reg}$ and the analytic expression for
$\Phi_{\rm reg}$ for a single
shifted Kerr hole of the same mass and angular momentum.  See text for
details.  }
\end{figure}

\begin{figure}
\epsfxsize=0.8\textwidth
\epsfbox{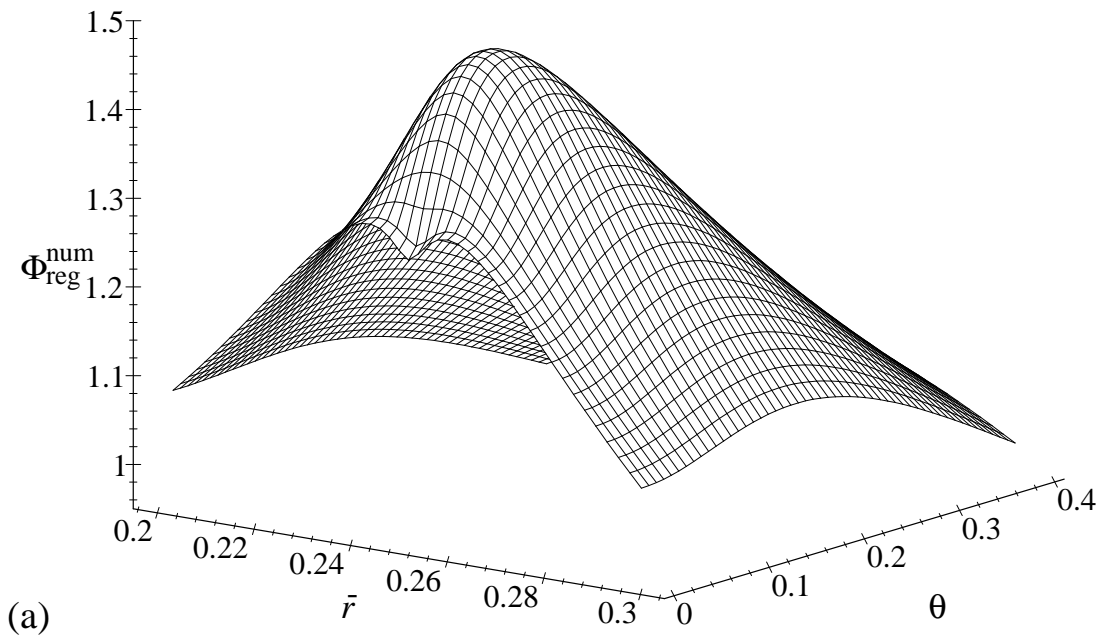}
\epsfxsize=0.8\textwidth
\epsfbox{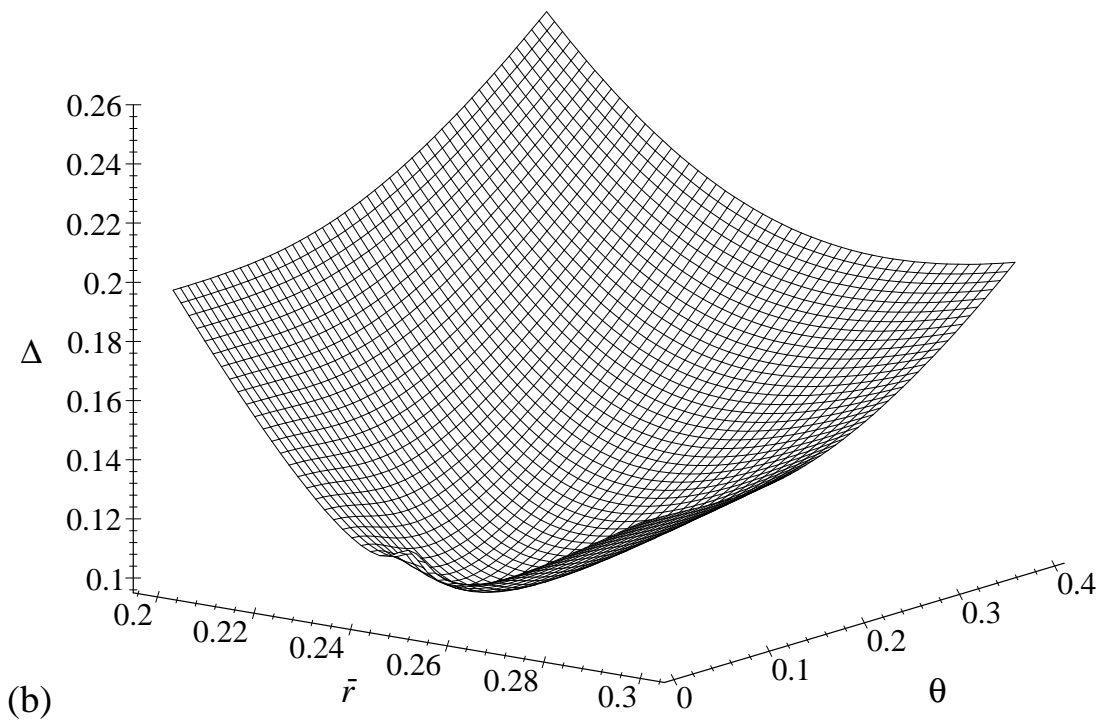}
\caption{\label{Figure-2}
Identical to Figure 2 except that $2 z_0=0.5$. At this 
smaller separation of the throats, the holes
lose local ``Kerricity''.
}
\end{figure}

In Fig.\ \ref{Figure-1}, part (a) shows $\Phi_{\rm reg}^{\rm num}$, the
computed regular part of $\Phi$, as a function of $\bar{r}$ and
$\theta$.  Part (b) shows $\Delta\equiv ({\Phi_{\rm reg}^{\rm num}}'
-\Phi_{\rm reg})/\Phi_{\rm reg}$, the relative difference between the
computational result and the analytic function $\Phi_{\rm reg}$, of
(\ref{Phi1}) for a single isolated hole of $M=0.5, J=0.225$. Since the
(two-throat) singular part for the two hole solution differs from the
(one-throat) singular part for an isolated solution, the difference
shown in (b) is {\em not} the same as the fractional difference
between the $\Phi_{\rm reg}^{\rm num}$ and $\Phi_{\rm reg}$, the 
regular part of
the shifted Kerr solution. Rather, part (b) indicates the size of
$({\Phi_{\rm reg}^{\rm num}}'-\Phi_{\rm reg})\equiv
\Phi_{\rm reg}^{\rm num}- \Phi_{\rm
reg}+\sqrt{M_2^2-a_2^2}/(2\bar{r}_2) $ 
with
$\bar{r}_2\equiv \sqrt{\bar{r}^2-2z_2\bar{r}\cos\theta+z_2^2}$.  In Fig.\ 
\ref{Figure-1} the separation parameter is $z_0=2.0$, and the computed
hole exhibits local ``Kerricity,'' i.e., the deviation from the Kerr
geometry is small, less than 4\% in the neighborhood of the throat.
Figure \ref{Figure-2} shows the analogous results for a separation
parameter $z_0=0.25$. Here, the deviation from the Kerr geometry is
significantly larger, on the order of 12\%. Yet
smaller values of $z_0$ produce yet larger deviations from the Kerr
geometry.

For definiteness we have described the axisymmetric superposition of
two equal mass, equal spin, Kerr holes, but it should be clear that
the method of superposition is more widely applicable to the
axisymmetric superposition of solutions. The general principle is to
put the initial value solutions to be added into the form
(\ref{genmet}). The individual metric $q$ functions and extrinsic
curvature $\widehat{H}$ functions are then superposed as in (\ref{supers}),
and the momentum constraint (\ref{momcon}) is guaranteed to be
satisfied. The last step is to choose the singularities, and
(numerically) to solve the Hamiltonian constraint for
$\Phi_{\rm reg}$. This can be done, for an arbitrary axisymmetric array of
Kerr holes, but the method is not limited to Kerr holes. Any solution
that can be put in the form (\ref{genmet}) can be superposed.  In
particular, the rotating holes of the Bowen-York\cite{BowenYork}
formalism can be used as building blocks. For these conformally flat
solutions, $q=0$, and superposing {\em only} Bowen-York holes will
lead to a conformally flat initial value solution; in this case our
method is simply a special case of the more general superposition that
is valid in the Bowen-York formalism. More interesting is the
possibility of superposing arbitrary mixtures of Bowen-York and Kerr
holes, or Kerr holes with distorted Kerr holes.

\vspace*{.2in}

\section*{III. Close limit solutions}
\begin{figure}
\epsfxsize=0.8\textwidth
\epsfbox{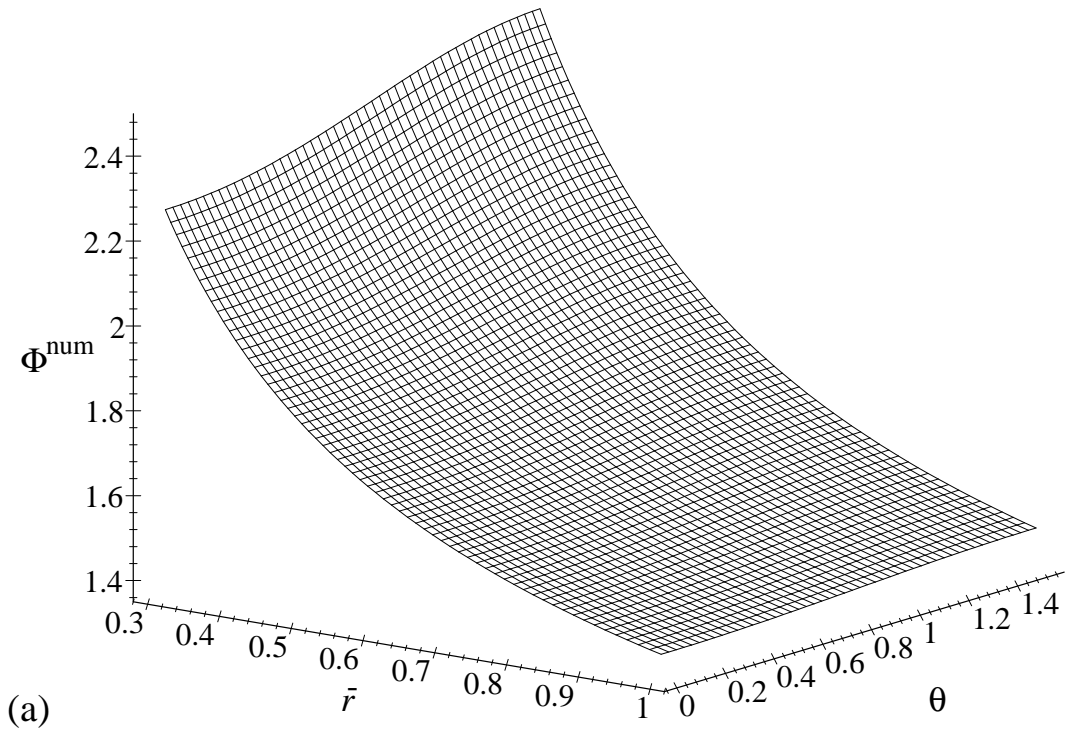}
\epsfxsize=0.8\textwidth
\epsfbox{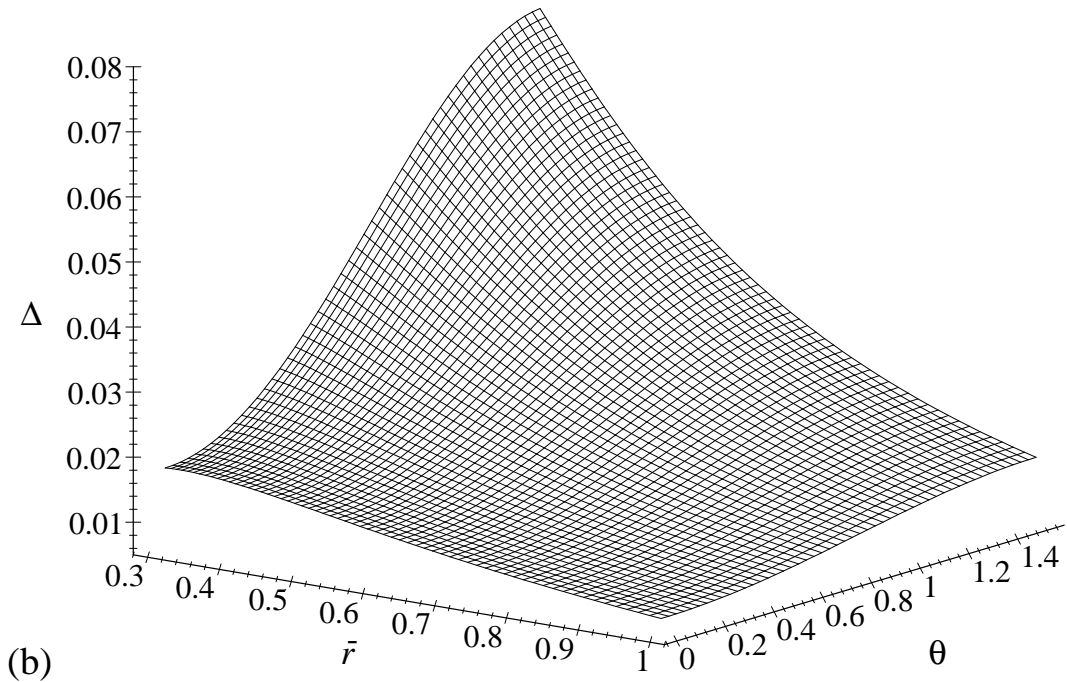}
\caption{\label{Figure-I}
  The conformal factor in the close limit.  Part (a) shows
  $\Phi^{\rm num}$, the conformal factor computed for two holes with equal
  masses $M_1=M_2=0.5$ and equal angular momenta $J_1 = J_2 =0.225$,
  placed symmetrically at $z_1=0.02$ and $z_2=-0.02$.  Part (b) shows
  the fractional difference $\Delta\equiv (\Phi^{\rm num}-\Phi_K)/\Phi_K$
  between the two hole solution and the conformal factor $\Phi_K$ for
  a Kerr hole of the same angular momentum and mass. See text for
  details.  }
\end{figure}

For solutions constructed by superposition as described above, the
close limit is not a single Kerr hole. In the two Kerr hole solution,
for example, if we let $|z_1-z_2|\rightarrow0$, the initial solution
will have an all encompassing horizon surrounding both the throat at
$z_1$ and at $z_2$. Outside this horizon the initial value solution
will {\em not} approach the Kerr solution. We illustrate this in
Fig.\ \ref{Figure-I}.  In part (a) we plot the computed
conformal factor $\Phi$ for the superposition of two identical holes,
with masses $M_1=M_2=0.5$ and angular momenta $J_1 = J_2 =0.225$,
placed at $z_1=-z_2=z_0=0.02$.  In part (b) we plot $\Delta$, 
the relative difference between the computed 
solution and a Kerr solution.
The comparison Kerr geometry is chosen to have
angular momentum $J=J_1+J_2=0.45$. For any other choice the extrinsic
Kerr geometry would not agree with that for the two hole solution at
large $\bar{r}$.  By looking at the large $\bar{r}$ behavior of the
computed $\Phi$ it was determined that the two hole solution had a
mass of $M=0.824$. (This is less than $M_1+M_2=1$ as should be
expected.)  If we were to compare our solution for $\Phi$ with a Kerr
hole of any other mass, there would be unacceptable differences at
large $\bar{r}$.

In Fig. \ref{Figure-I}, the left (small radius) boundary of the
surface is at the approximate location of the initial horizon.  As one
would expect, the plot in (b) shows that the maximum deviation from
the Kerr geometry (around 8\%) occurs on the horizon. The crucial
feature of these plots is that {\em they do not change significantly 
  as $z_0\rightarrow 0$.} The close limit of the two holes is not a
single hole; the deviation from the Kerr geometry exhibited in part
(b) does not decrease with decreasing separation. This is in clear
contrast to the nonrotating case, in which the deviation outside the
horizon, decreases monotonically with the separation of the holes.

There are two very different reasons for wanting a solution with a
single-Kerr close limit. The first is that the evolution of a binary
system will result in a single Kerr hole as the evolved close limit,
so we might want an initial value solution with this feature. (This
will be discussed further in the next section.) Of more practical
interest, if we are to do close-limit perturbation theory, we must
have a simple stationary solution as the close limit.  

It is, in fact, very simple to construct a superposition with the
desired close limit. We can take the metric function $q$ and the
conformally related extrinsic curvature functions $\widehat{H}$ to be
precisely those of a single Kerr hole of mass and spin $M,a$, i.e.,
the functions given by (\ref{qKerr}) --- (\ref{HFKerr}) and
(\ref{rdef}). We could then try to impose the topology of two throats
by taking $\Phi_{\rm sing}$ to have the two-throat form of
(\ref{BLfor2holes}). But in this case, the righthand side of
(\ref{hamforphireg}) would have a $1/\bar{r}^6$ singularity at
$\bar{r}\rightarrow0$, if $\Phi_{\rm sing}+\Phi_{\rm reg}$ were 
finite there, since
$\widehat{H}_{EK}$ does not vanish at $\bar{r}\rightarrow0$.  In fact, the
``nonsingular'' function $\Phi_{\rm reg}$ we compute diverges as
$1/\sqrt{\bar{r}}$ as $\bar{r}\rightarrow0$. With the resulting
$\Phi^4\sim\bar{r}^{-2}$ behavior of the conformal factor, the topology
of the metric is ${\cal R}^1\times{\cal S}^2$ near
$\bar{r}\rightarrow0$. It is possible in principle for us to make
$\Phi_{\rm reg}$ truly regular by adding a term of the form
$h(\theta)/\sqrt{\bar{r}}$ to $\Phi_{\rm sing}$. The angular function
$h(\theta)$ is the solution of the nonlinear ordinary differential
equation that results when the Kerr forms of $\widehat{H}_E,
\widehat{H}_F$ and
$q$, and the {\em ansatz} $\Phi=h(\theta)/\sqrt{\bar{r}}$ are put into
(\ref{hamconst}), and the $\bar{r}\rightarrow0$ limit is taken. Since
this is numerically inconvenient, we have chosen to keep the
$1/\sqrt{\bar{r}}$ singularity as a part of $\Phi_{\rm reg}$, and to use the
boundary condition for (\ref{hamforphireg}) that
$\sqrt{\bar{r}}\Phi_{\rm reg}$ is regular at $\bar{r}\rightarrow0$.

An alternative is for us to choose to add a term of the form
$\kappa/\bar{r}$ to $\Phi_{\rm sing}$. For any value of $\kappa$ this will
cause the extrinsic curvature term on the righthand side of
(\ref{hamforphireg}) to vanish at $\bar{r}\rightarrow0$. Of course, by
adding such a term to $\Phi_{\rm sing}$, we are adding another
asymptotically flat region, i.e., another throat. Our solution, then
can be taken to represent an ``extra'' black hole. We could minimize the
intrusiveness of this unwanted additional throat by choosing $\kappa$,
the parameter that governs its size, to be arbitrary small. In this
case, at small $\bar{r}$ the extrinsic curvature term in
(\ref{hamforphireg}) will become very large, of order $J^2/\kappa^6$
around $\bar{r}\sim\kappa$.

If we are to use the above choices of topology, either the
$\Phi\sim1/\sqrt{\bar{r}}$, or $\Phi\sim1/\bar{r}$, at
$\bar{r}\rightarrow0$, we no longer have the two throat topology that
represents two holes. But if this superposition is to be made only in
the case the two holes are quite close, the additional central
topological feature will be deep inside an all-encompassing horizon,
and will be irrelevant to close-limit analysis or for numerical
relativity if horizon excision is used. As we shall argue further in
the next section, this means that the unwanted central topological
feature in no way makes our close-limit superposition less useful.  In
this section we give numerical results for close-limit superpositions,
and further details of how they can be used for numerical and analytic
work.

In our close limit superposition to form a single Kerr hole of mass
$M$ and spin $a$, in addition to choosing the Kerr forms for
$q,\widehat{H}_E$ and $\widehat{H}_F$, we must also choose 
$\Phi_{\rm sing}$ to have
the form of (\ref{BLfor2holes}), with the numerators in the second and
third terms replaced by $1/2\,\sqrt{M^2-a^2}$.  It is only for this
choice of singular structure that the mass contained in $\Phi_{\rm sing}$,
combined with the mass coming from the solution of
(\ref{hamforphireg}), will be the mass $M$ in the $z_0\rightarrow0$
limit.

\begin{figure}
\epsfxsize=0.8\textwidth
\epsfbox{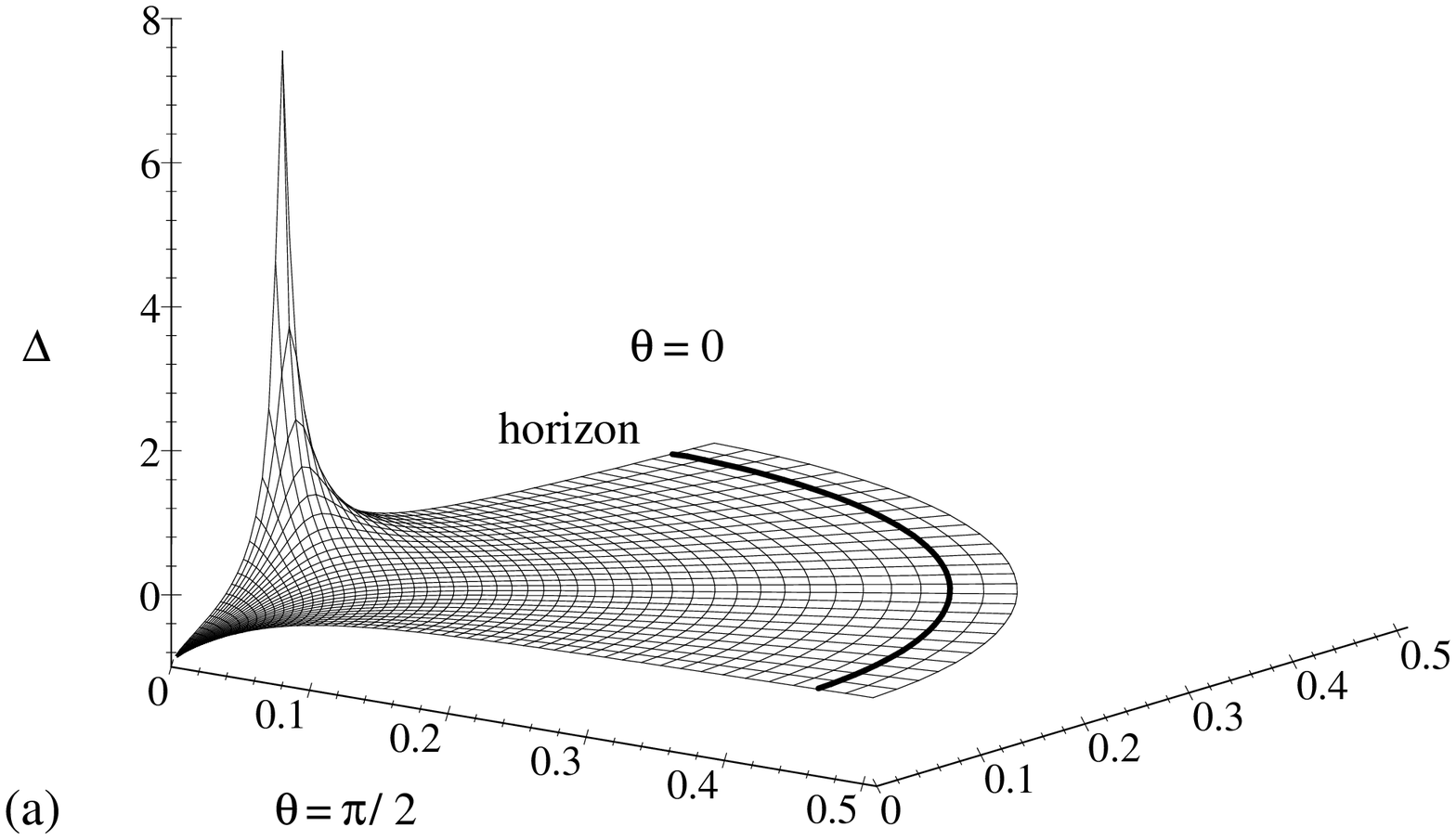}
\epsfxsize=0.8\textwidth
\epsfbox{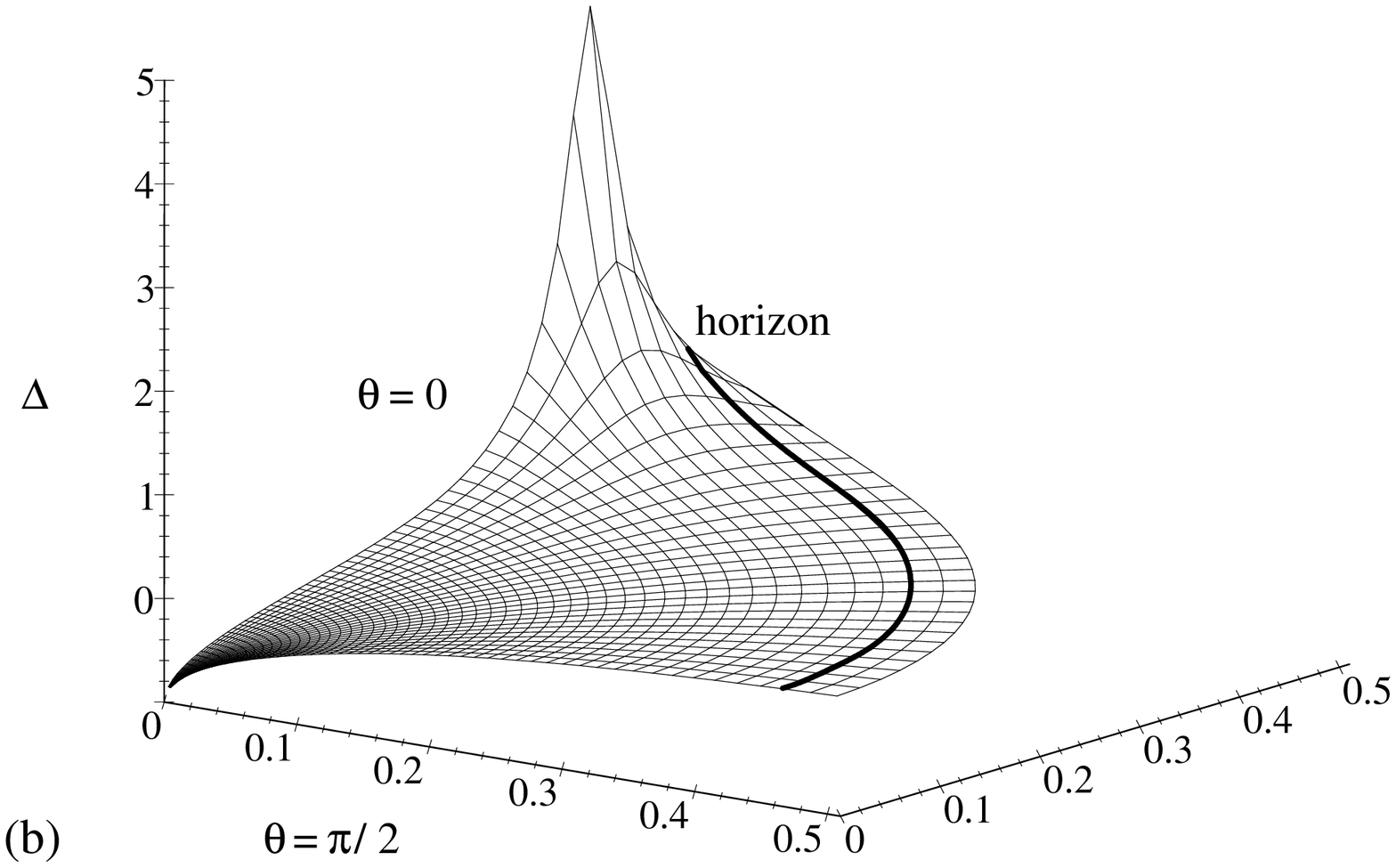}
\caption{\label{Figure-II}
  The fractional difference $\Delta\equiv (\Phi^{\rm num}-\Phi_K )/\Phi_K$
  between $\Phi^{\rm num}$, the numerically computed conformal factor for
  $M=1.0, J=0.45$ and $\Phi_K$, the conformal factor for a Kerr hole
  of the same mass and angular momentum.  In (a), the separation is
  given by $2 z_0=0.2$, and the difference is less than $8\% $ outside
  the horizon (shown as a thick curve).  In (b), $2 z_0=0.8$, and
  there is a sizeable difference between the close-limit and the Kerr
  conformal factors.  }
\end{figure}

In Fig.\ \ref{Figure-II} we illustrate the solution to
 (\ref{hamforphireg})  for the 
problem just described.  For two close holes, the plots shows the relative
difference between the numerically computed $\Phi$ and $\Phi$ for a
single Kerr solution with the same mass
and angular momentum. The initial location of the horizon
is indicated.  These plots demonstrate that outside the horizon $\Phi$
approaches the Kerr form as the separation between the throats
(i.e., between the singularities in $\Phi_{\rm sing}$) is reduced. The 
solutions deep inside the horizon are very different, as they must be, 
no matter how small the separation.
As explained above, for these choices there will be a formal
singularity at $\bar{r}z$ corresponding to a ${\cal R}^1\times{\cal S}^2$
feature. 

So far we have presented initial data solutions in this section which
can be used for close-limit analytic work, and in Sec.\ II we
presented solutions that accurately describe two Kerr holes in the far
separation limit. In the remainder of this section we will show that, 
with very little additional complexity, these close and far limit 
solutions can be combined. To simplify the description of how this is 
done, we will confine attention to the case of two
identical holes, of parameter $M_1=M_2\equiv M/2$ and angular momentum
$J_1=J_2\equiv J/2$, symmetrically placed at $z_1=-z_2\equiv z_0$, as
in the examples of Sec.\ I. 

To combine solutions we need to introduce a mixing function $f(z_0/M)$
that is used to weigh the amount of the far-limit solution that is to
be mixed with the close limit solution.  For the choice made for
$J_1/M_1$, let us take $z_{crit}$ to be the largest value of $z_0$ for
which the initial horizon encloses all throats (the two throats
representing the holes, and the feature at $\bar{r}=0$). Then the mixing
function must have the property that $f=1$ for $z_0/M> z_{crit}/M$. It
must also have the close-limit property that $f\rightarrow0$ for
$z_0/M\rightarrow0$.  The mixing function might be taken, for
example, to be
\begin{equation}
  \label{fposs}
  f\equiv\left\{
    \begin{array}[c]{ll}
1&\mbox{for $z_0>z_{crit}$}\\
1-\left(z_0-z_{crit}\right)^2/z_{crit}^2&\mbox{for $z_0\le z_{crit}$}
    \end{array}
\right.\ .
\end{equation}
For our mixed solutions, we take $q$ and the $\widehat{H}$ functions to be
\begin{eqnarray}
  q&=&f[q_1(M_1,J_1)+q_2(M_2,J_2)]+(1-f)\,q_K(M,J)\; ,\label{qcomb}\\
\widehat{H}_E&=&f[\widehat{H}_{E1}(                  M_1,J_1) 
+\widehat{H}_{E2}(M_2,J_2)]\nonumber\\
&+&(1-f)\,\widehat{H}_{EK}(M,J)\; ,\label{Hcomb1}\\
\widehat{H}_F&=&f[\widehat{H}_{F1}(M_1,J_1)
+\widehat{H}_{F2}(M_2,J_2)]\nonumber\\
&+&(1-f)\,\widehat{H}_{FK}(M,J)\ \label{Hcomb2}.
\end{eqnarray}
For clarity we have written the expressions in (\ref{qcomb}) --
(\ref{Hcomb2}) with parameters $M_1, J_1,M_2, J_2,M,J$, and have not
invoked the special values of $M_1,M_2,J_1,J_2$.  The functions ($q_1,
\widehat{H}_{E1}, \widehat{H}_{F1}$), with subscript 1, are the 
shifted solutions
of (\ref{2q1}), (\ref{HEtrans_fnl}), and (\ref{HFtrans_fnl}), and
similarly for the functions with subscript 2.  The functions with $K$
subscript are the ``pure'' Kerr functions from
(\ref{qKerr})--(\ref{HFKerr}).  For the mixed solution we take the
function $\Phi_{\rm sing}$ to have the form
\begin{equation}      
  \label{3throat}
  \Phi_{\rm sing}=
1+        \frac{1}{2}\, 
      \frac{f\sqrt{M_1^2-a_1^2}
+(1-f)\sqrt{M^2-a^2}/2}{\sqrt{\bar{r}^2-2z_1\bar{r}\cos\theta+z_1^2}}
+ \frac{1}{2}\frac{f\sqrt{M_2^2-a_2^2}+(1-f)\sqrt{M^2-a^2}/2}
        {\sqrt{\bar{r}^2-2z_2\bar{r}\cos\theta+z_2^2}}
  \, .
\end{equation}

For separations large enough so that a horizon does not enclose both
holes, the solution, is identical to the solution of Sec.\ I, and
therefore has the correct far limit, i.e., in the limit of large
separation the solution near each throat approaches that of an
isolated Kerr throat. For $z_0/M\ll1$, the mixing function $f$
approaches 0, and the solution approaches the 
Kerr solution, and hence the mixed solution must 
have the correct close-limit, i.e., in the limit of small separation the
solution outside the horizon approaches that of a single Kerr hole.

It is interesting to note that our sequence of solutions, with mass
and the angular momentum the same at large and at small separations,
cannot produce a rapidly rotating final hole; for extreme individual
holes, with $M_1=J_1$,  the final hole has $a=J/M^2=0.5$. If we
want a close limit final hole that is rapidly rotating we cannot use a
constant-mass, constant-angular momentum sequence. This reminds us
that simply adding rotating holes does not automatically produce
something of physical relevance, or at least, does not produce what we
may want.  This is a point we will be coming back to in the next
section.

\section*{IV. Summary and conclusions}
We have presented a general method for constructing axially symmetric
superpositions of Kerr black holes. The crucial feature of this method
is the adoption of a 3-metric form, that of (\ref{genmet}), with
which the momentum constraint for the conformally related extrinsic
curvature reduces to a linear equation, independent of the 3-geometry.
In this limited sense, the approach has the simplicity of the
Bowen-York\cite{BowenYork} conformally flat approach.

In the case of sizeable separation between the superposed holes, the
method produces holes with ``local Kerricity,'' and has the advantage
of great simplicity and easy numerical implementation. The method has
been illustrated with numerical results for two equivalent (equal
mass, equal spin) holes, but many extensions are possible. It should
be clear how to superpose any number of holes with ``local
Kerricity,'' or holes without local Kerricity, in particular locally
Bowen-York holes.

A crucial feature of our superposed solutions is that their
close-limit is {\em not} a single Kerr hole outside the horizon. We
have argued that unlike the spherically symmetric case, it should be
expected on physical grounds that two locally Kerr holes, no matter
how close they are, will not superpose into a single Kerr hole. Though
the ``failure'' of the close limit is physically correct, it is
inconvenient. We would like to do close limit perturbation theory,
which will at the very least be of value for checking numerical codes
for colliding rotating holes. We wish to emphasize that the {\em
correct} close limit data for two holes, is the data produced by the
evolution of a pair of holes falling into each other. A guess at what
that initial data solution is, has little chance of being accurate.
This was true, of course, also for initial data solutions representing
nonrotating black holes, but in that case the variety of choices was
small, and the radiation evolving from the close limit initial data
seemed to be rather insensitive to the choice of details. This is very
unlikely to be the case for rotating holes; here the richness of the
geometry and extrinsic curvature gives a much greater scope of
choices.  

For the reasons just given, our focus is on close limit calculations
as a check on numerical relativity calculations, so definitiveness and
simple implementation are of importance. We have given, in Sec.\,III,
an alternative superposition scheme, which is very easily implemented,
and which has the correct close limit, but it has a singularity at the
midpoint between two equal holes. That singularity will be hidden deep
inside the horizon of a close limit solution, and is of concern only
for numerical relativity methods that require a solution throughout
the interior of the horizon. The presence of the singularity does not
in any way suggest that the solution outside the horizon is a good or
bad representation of the late stage of infall of two Kerr holes, or
whether the radiation produced outside the horizon is for any reason a
good or bad test of numerical relativity.

Having the singularity hidden inside the horizon means of course, that
this type of initial data is only applicable in the close limit. Since
it may be useful to have a sequence of solutions which includes both
large and small separations, we have also given, in Sec.\,III, a
method of mixing both types of superpositions so that for large
separations one has locally Kerr holes and no spurious singularities, 
and in the close limit one approaches a single Kerr hole, and has a
spurious singularity hidden inside the horizon.

We intend to use the close limit, and mixed-sequence, initial data
solutions as a starting point for close limit computations of
radiation.  We have already converted the superposition solutions, in
the close limit, to Cauchy data for the Teukolsky equation\cite{teuk}
and will be evolving a number of examples. With the technical tools in
place to do studies of evolution to a rapidly rotating hole, we will
be able to answer a number of interesting questions, such as the range
of validity of the slow rotation limit.

\begin{acknowledgements}
We thank John Baker and Raymond Puzio for helpful suggestions. This work
was partially supported by NSF grant PHY-9507719.
\end{acknowledgements}

\end{document}